\documentclass[12pt]{article}
\usepackage{epsfig}

\begin{document}

\begin{titlepage}

\setlength{\textheight}{23.5cm}

\begin{center}
{\Large \bf Multiparton scatterings at the LHC} \vspace{1cm} \\
        { \bf Alessio Del Fabbro}\footnote{E-mail:{\it delfabbr@ts.infn.it}}
    and
    { \bf Daniele Treleani}\footnote{E-mail:{\it daniel@ts.infn.it}}
\vspace*{.5cm}  \\
        {\it Dipartimento di Fisica Teorica, Universit\`a di Trieste, \\
        Strada Costiera 11, I-34014 Trieste}  \\
    and \\
    {\it INFN, Sezione di Trieste  \\
        via Valerio 2, I-34127 Trieste }
\vspace{.4cm} \\
\end{center}
\vspace{1cm}
\begin{abstract}
The large parton flux at high energy gives rise to events where
different pairs of partons interact contemporarily with large
momentum exchange. A main effect of multiple parton interactions
is to generate events with many jets at relatively large
transverse momenta. The large value of the heavy quarks production
cross section may however give also rise a sizable rate of events
with several $b$-quarks produced. We summarize the main features
of multiparton interactions and make some estimate of the
inclusive cross section to produce two $b{\bar b}$ pairs within
the acceptance of the ALICE detector.
\end{abstract}

\begin{flushbottom}
\begin{footnotesize}

\end{footnotesize}
\end{flushbottom}

\end{titlepage}

\section{Multiparton scatterings and Jets in proton-proton
collisions}

The description of jet production in hadronic collisions with a
momentum transfer of the order of the c.m. energy, where the
process is characterized by a single large energy scale factor,
represents one of the most successful applications of the
conventional, fixed$-x$, perturbative QCD. In this regime higher
orders in the perturbative expansion of the elementary interaction
are related directly to processes where an increasing number of
large $p_t$ jets are produced and virtual corrections are a
rapidly converging expansion in powers of the strong coupling
constant. The simplest perturbative scheme is however no more
adequate when two or more different scales become relevant. An
interesting case where two scales play an important role is when
the c.m. energy of the partonic process, although much larger than
the hadronic scale, is nevertheless very small as compared with
the c.m. energy of the hadronic interaction, namely $\hat{s}\ll
s$, where ${\hat s}\equiv xx's$, with $x$, $x'$ the momentum
fractions of the interacting partons and $s$ the Mandelstam
invariant of the hadronic collision. In this kinematical regime
the typical final state is characterized by the presence of many
jets and, while the partonic collision is still well described by
the conventional perturbative approach, the overall semi-hard
component of the hadronic interaction acquires a much richer
structure. A recent analysis performed by
CDF\cite{Affolder:2002zg} of the jet evolution and underlying
event in $p\bar{p}$ collisions shows that, while the leading jet
is fairly well described by QCD Monte Carlo models in a wide
kinematical range, the models fail do describe correctly the
underlying event, which is populated by several further jets at a
relatively low $p_t$. From the theoretical point of view one may
get in touch with the problem looking at the cross section to
produce large $p_t$ jets, integrated on the exchanged momentum
with the cutoff $p_{cut}$. When $p_{cut}$ is moved towards
relatively small values, the integrated cross section, being
divergent for $p_{cut}\to0$, becomes very large already well
inside the kinematical regime where perturbation theory is
meaningful, so that one faces a unitarity problem. Indeed the
cross section for production of jets with $p_t>5$GeV measured by
UA1\cite{Arnison:1986vk} is about $15-20mb$, in $p{\bar p}$
collisions at a c.m. energy of $900$ GeV.
\begin{figure}
\begin{center}
\epsfig{figure=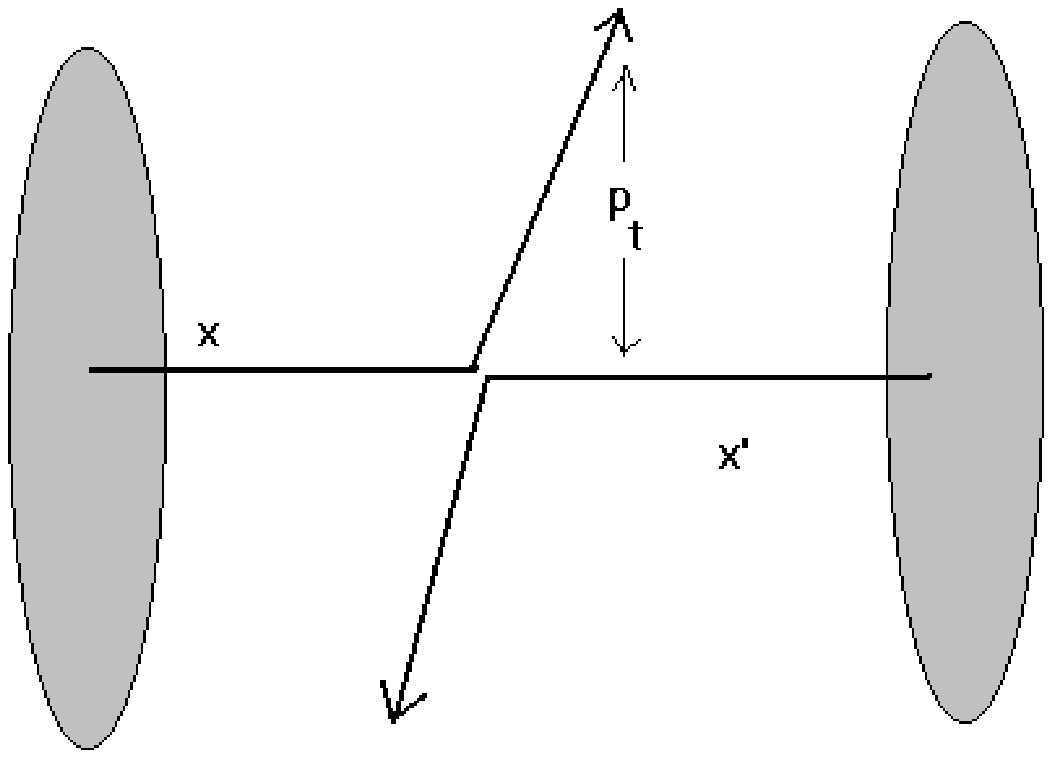,height=3in} \caption{Single parton
scattering} \label{single} \vskip.3in
\epsfig{figure=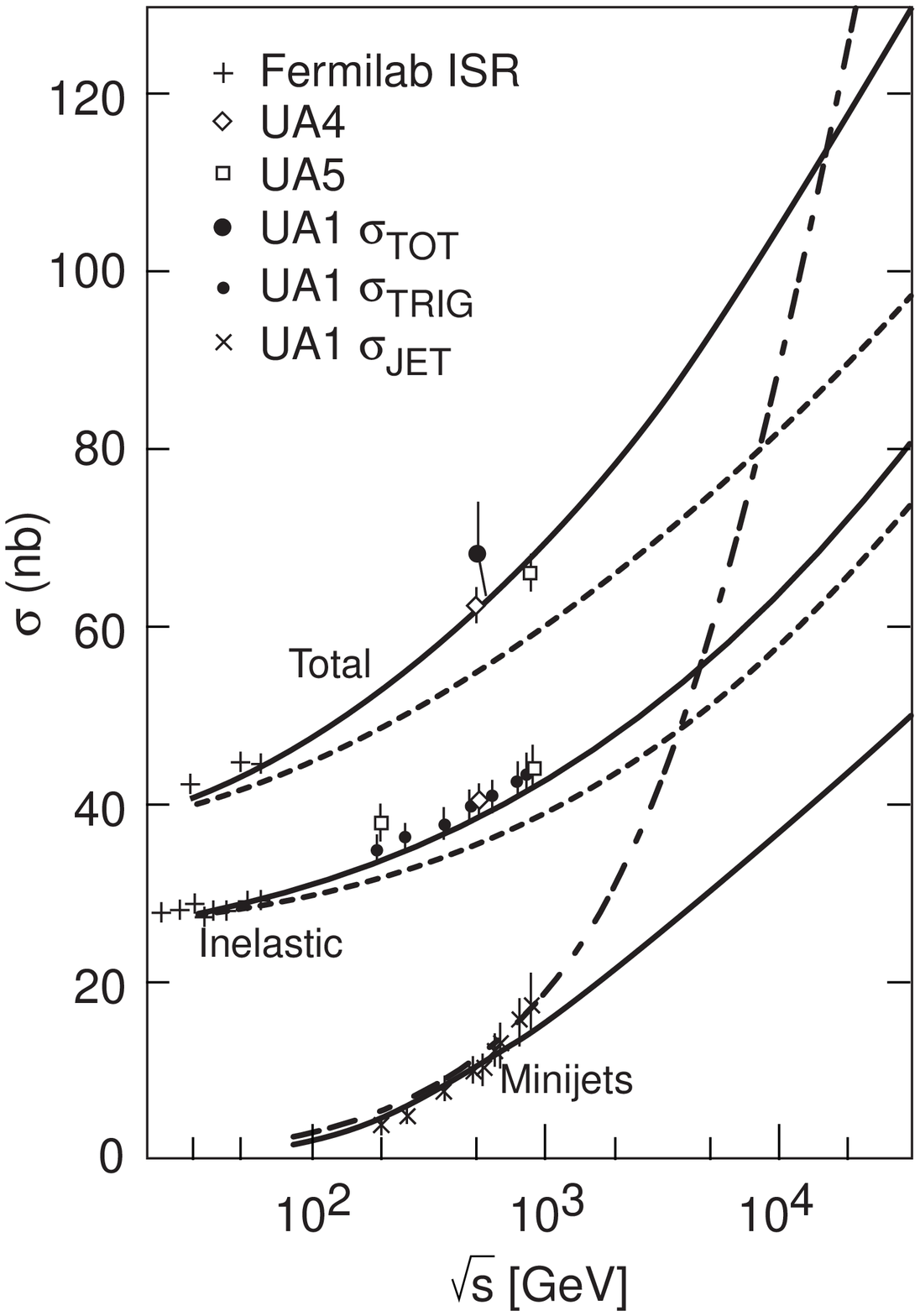,height=4.5in} \caption{Total, inelastic
nondiffractive and minijet cross sections as a function of the
c.m. energy[3]} \label{capella}
\end{center}
\end{figure}
In fig.\ref{capella} the total, the inelastic non-diffractive and
the cross section to produce jets with $p_t>5$GeV (minijets) are
plotted as a function of energy\cite{Capella:1986cm}, showing that
one foresees a high energy regime where the integrated inclusive
cross section of jet production (long-dashed curve in the figure)
exceeds the value of the total inelastic cross section. If
considering hadron-nucleus and nucleus-nucleus collisions, the
unitarity problem in jet production appears at larger transverse
momenta and with smaller nucleon-nucleon c.m. energies.

The origin of the unitarity problem is in the large flux of
partons, active for producing large $p_t$ processes at high
energies. When the incoming partons flux is very large there is a
finite probability of producing events where different pairs of
partons interact independently in a given hadronic interaction,
with momentum exchange above $p_{cut}$, generating states with
many jets at a relatively large $p_t$. In the integrated inclusive
cross section of jet production, those events count with the
multiplicity factor of the partonic collisions and, as a
consequence, the inclusive cross section is not bounded by the
value of the total inelastic cross section. In the kinematical
region where the inclusive cross sections of jet production is
comparable to the total inelastic cross section, the average
number of partonic interactions in a typical inelastic event is
close to one and the probability of multiple parton collisions is
large. Interestingly in this regime the single scattering
expression of perturbative QCD is still meaningful, and represents
correctly the integrated inclusive cross section, which, as
required by the cancellation property discussed by Abramovsky,
Gribov and Kancheli\cite{Abramovsky:fm} for interactions where
multiple exchanges take place, is given by the single scattering
term expression. Notice that, since in the inclusive each partonic
collision contributes with its multiplicity, the inclusive cross
section is proportional to the average number of multiple
interactions, which ultimately is also the reason of the simple
theoretical result of the AGK cancellation. On the other hand,
precisely for the same motivation, the inclusive cross section
gives only a rather limited information on the production
dynamics.

To gain a deeper insight one needs to look at different and less
inclusive physical observables. From the experimental point of
view the simplest quantity of this kind is the semi-hard cross
section, namely the cross section which counts all events with at
least one jet produced, which is related to the probability of not
having any hard interactions at all in a inelastic event. The
semi-hard cross section is hence bounded by the value of the total
inelastic cross section and therefore characterized by an
intrinsic dimensional scale factor. A consequence is that the
semi-hard cross section, contrarily to the inclusive cross
section, cannot be obtained by evaluating the single scattering
expression of perturbative QCD (where the only relevant
dimensional factors are the kinematical variables). In fact the
semi-hard cross section needs an infinite number of perturbative
terms to be evaluated, corresponding to all possible processes
with any number of multiple parton collisions. To obtain a
complete information on the semi-hard interaction dynamics one
needs to measure systematically all the moments of the
distribution of multiple parton collisions, which implies
measuring all inclusive cross sections with any number of produced
jets and isolating the corresponding multiple parton scattering
contributions. A given configuration with many jets may be in fact
originated in various ways with different elementary partonic
processes. Each multiple scattering term is nevertheless
characterized by a different dimensional scale factor so each
multiparton scattering process has a peculiar dependence on the
c.m. energy and on the lower cutoff in $p_t$, while the typical
pattern of the final state produced is obtained by superposing
single scattering events. The different contributions to the
semi-hard cross section, with a given number of elementary
partonic collisions, may be therefore disentangled by looking at
the energy and cutoff dependence of the cross section and by
studying the pattern of the final states produced. A possible way
to obtain the dimensional scale factors characterizing the
different multiparton scatterings, is to measure the relative
rates of production of 1, 2, 3 etc. jets as a function of the
lower cutoff in $p_t$. By moving from the large to a relatively
low $p_t$ region the ratios in fact change, as compared to the
expectations bases on the conventional single scattering
production mechanism, because of the increasingly large
contribution of multiple scatterings at relatively low $p_t$. The
measurement of the azimuthal correlations between charged tracks
at relatively large $p_t$ is a further tool to identify production
processes characterized by independent hard collisions at the
parton level.

The non-perturbative input to a multiparton interaction is a
multiparton distribution, which contains independent informations
on the hadron structure, as compared to the one-body parton
distributions usually considered in large-$p_t$ processes.
Multiparton distributions depend in fact linearly on the
multiparton correlations of the hadron structure. Notice that
multi-parton scatterings have been already implemented in some QCD
Monte Carlo program, to the purpose of describing the minimum bias
event in high energy hadronic collisions\cite{Sjostrand:1987su}. A
basic hypothesis there is however that all input multi-parton
distributions are trivial (namely poissonians at a fixed impact
parameter), which amounts neglecting all multiparton correlations.
The difficulties of the QCD Monte Carlo programs to describe the
underlying event in hard collisions at
$1.8$TeV\cite{Affolder:2002zg}, even after switching on
multi-parton scatterings, and the results of the analysis of the
simplest case of multi-parton scattering, the double parton
collision\cite{Abe:1997bp}\cite{Abe:1997xk}, point on the contrary
in the direction of the existence of non trivial correlations
effects in the hadron structure. An exhaustive description of the
hard component of the interaction, in high energy hadronic
collisions, requires therefore an explicit investigation of each
different multiparton scattering process and the measurement of
all various correlation parameters.

\section{ Double parton scattering}

The simplest case of multiparton interaction is the double
collision. Correspondingly the simplest state produced is
represented by four jets with transverse momenta above the cutoff
$p_{cut}$ and balanced in pairs. The same state may also be
produced by the leading QCD $2\to4$ process. While the latter
process can be estimated at the three level without much
effort\cite{Berends:1989hf}\cite{Stelzer:1994ta}, the evaluation
of the double parton scattering contribution is in principle much
more uncertain. The main reason of uncertainty is represented by
the non-perturbative input needed to the double parton collision,
the two-body parton distributions, which contains the information
on the two-body parton correlations. All discussions on the double
parton collision process rely on assumptions on the two-body
parton correlations. Since multiple parton interactions are a
sizable effect at low $x$, where the parton flux is large,
correlations in momentum fraction should not be a major effect, so
that a reasonable hypothesis is to neglect correlations in
fractional momenta in the multi-parton distributions. Not all
possible correlations can however be neglected, since the
interacting partons must be correlated in transverse space,
belonging to the same hadron. It seems therefore reasonable, at
least in the low $x$ region, to make the hypothesis that the only
relevant correlations are those in the transverse parton
coordinates. With this assumption, on rather general grounds, the
inclusive double parton scattering cross section is given by the
product of two single parton scattering cross sections with a
scale factor, which depends smoothly on the c.m. energy and on the
cut-off and which, in general, depends on the parton process
considered. One might in fact expect different kinds of partons to
have a different distribution in transverse space inside the
hadron. The double scattering cross section $\sigma_D(A,B)$ for
the two parton processes $A$ and $B$ is then written as
\begin{equation}
\sigma_D(A,B)=\frac{m}{2}\sum_{ijkl}\Theta^{ij}_{kl}\sigma_{ij}(A)\sigma_{kl}(B)
\label{sigmad1}
\end{equation}
where $\sigma_{ij}(A)$ is the hadronic inclusive cross section for
two partons of kind $i$ and $j$ to undergo the hard interaction
$A$, while the partons $k$ and $l$ undergo the hard interaction
$B$ with cross section $\sigma_{kl}(B)$. The factor $m$ equals 1
when the two parton processes $A$ and $B$ are identical, while its
value is $2$ if they are distinguishable.
\begin{figure}
\begin{center}
\epsfig{figure=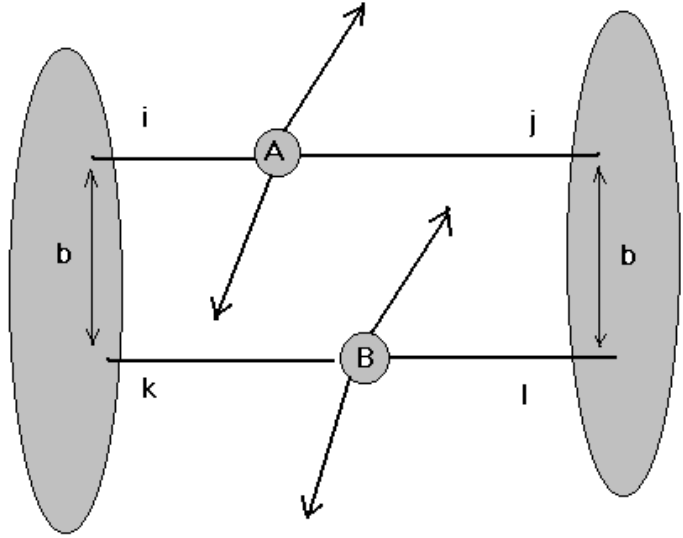,height=2.8in} \caption{Double parton
scattering}
\end{center}
\end{figure}
The factors
\begin{equation}
\Theta^{ij}_{kl}=\int d^2bF_k^i(b){F_l^j}(b) \label{fb}
\end{equation}
are geometrical coefficients, with dimensions of inverse cross
section. The function $F_k^i(b)$ represents the density of the
pair of partons $k,i$ in a hadron as a function of the relative
transverse distance $b$, normalized to one. Since the partonic
interactions $A$ and $B$, being characterized by a scale of
momentum transfer much larger than the hadronic dimension ($p_t\gg
1/R$) are well localized in transverse space inside the hadron,
the relative transverse distance between the pair $k,i$ and
between the pair $l,j$ of the two colliding hadrons must be
equal, in order to have the alignment necessary for the double
interaction to take place. The cross section is then obtained by
summing over all possible configurations, with the interacting
parton pairs at the same relative transverse distance $b$ in the
two hadrons\cite{Paver:1982yp}.

Double parton collisions have been measured by
CDF\cite{Abe:1997bp}\cite{Abe:1997xk} looking at final states with
three minijets and one photon. The cross section has been
expressed as
\begin{equation}
\sigma_D=\frac{1}{\sigma_{eff}}\quad\sigma_S(\gamma j)\sigma_S(jj)
\label{cdf}
\end{equation}
which may be obtained from Eq.\ref{sigmad1} making the assumption
that the geometrical coefficients $\Theta^{ij}_{kl}$ do not depend
on the different kinds of interacting partons $ijkl$. The
experiment gives as a output the value of the scale factor:
\begin{equation}
\sigma_{eff}=14.5\pm1.7^{+1.7}_{-2.3} \quad{\rm mb}\label{seff}
\end{equation}
Given the kinematical range available to the CDF experiment, the
sum on the different kinds of partons is dominated by the
contribution where three of the indices in Eq.\ref{sigmad1} refer
to gluons and one to quarks, so that basically
$\sigma_{eff}\simeq1/\Theta^{qg}_{gg}$.

While, at least in the limited kinematical range available, the
scale factor does not show evidence of dependence on the
fractional momenta of the interacting partons, consistently with
the hypothesis of no correlation in the $x$-variables, the actual
measured value of $\sigma_{eff}$ points, on the contrary, towards
the existence of non-trivial correlations of the proton structure
in transverse space\cite{Calucci:1997ii}\cite{Calucci:1999yz}. In
the case of no correlations, one would in fact expect
$\sigma_{eff}=h\sigma_{NSD}$, where $h$ is a geometrical
enhancement factor and $\sigma_{NSD}$ the nucleon
non-single-diffractive cross section (namely the inelastic cross
section minus the single diffractive cross section) whose value,
measured by CDF, is $50.9\pm1.5$ mb. The geometrical enhancement
factor keeps into account that multiple parton collisions are more
likely to take place at small impact parameters, where the overlap
of the matter distributions of the two interacting hadrons is
larger. The indication obtained, considering different
possibilities for the density of partons in space, is that the
value of $h$ should be close to $1/2$, which would give a
$\sigma_{eff}$ roughly twice as big as the measured value. One may
then argue that the experimental indication points towards the
existence of non trivial correlation effects in transverse space.

A natural implication of the existence of correlations in
transverse space is that different pairs of partons are
characterized by different typical transverse distances, which
amounts to a non negligible dependence of the scale factors
$\Theta^{ij}_{kl}$ on the different indices. Even if this is the
case, the double parton scattering cross section may still be
expressed by the factorized form in Eq.\ref{cdf} used by the CDF
experimental analysis. When $\Theta^{ij}_{kl}$ has a sizable
depence on the indices, $\sigma_{eff}$ however not only changes
with the kind of double parton process considered, but also
depends on the c.m. energy of the hadronic collision and on the
phase space cuts applied, since, given a definite final state, the
luminosity of the different kinds of initial state partons
contributing to the process changes both with energy and with the
cuts applied to the final state\cite{DelFabbro:2000ds}. On the
other hand the indication of non-trivial correlations in
transverse space implies a structure of the proton much richer
than obtainable through deep inelastic scattering experiments with
electron beams, where the information accessible is limited by the
pointlike structure of the projectile.
\begin{figure}
\begin{center}
\epsfig{figure=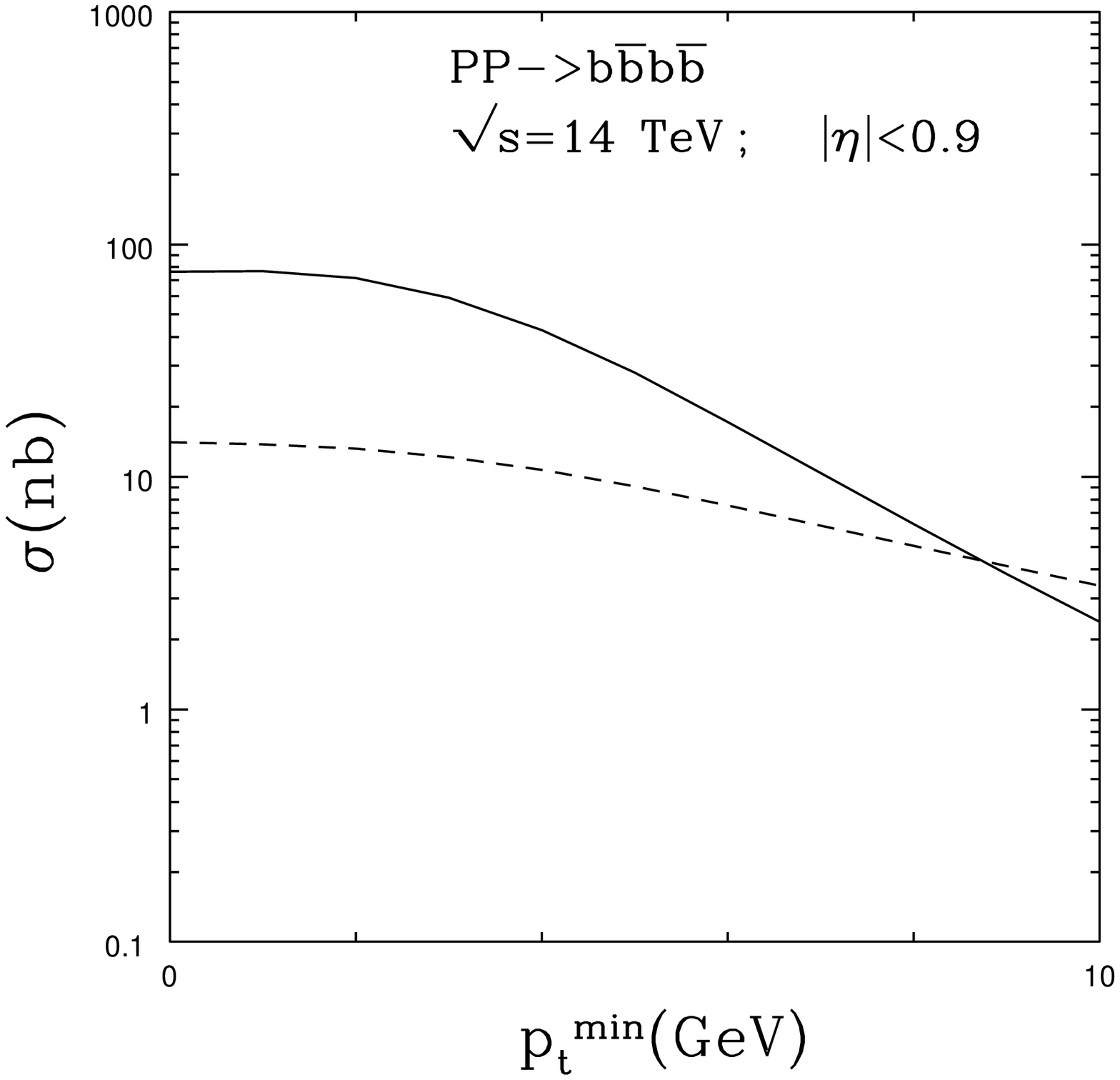,height=3.1in}
\epsfig{figure=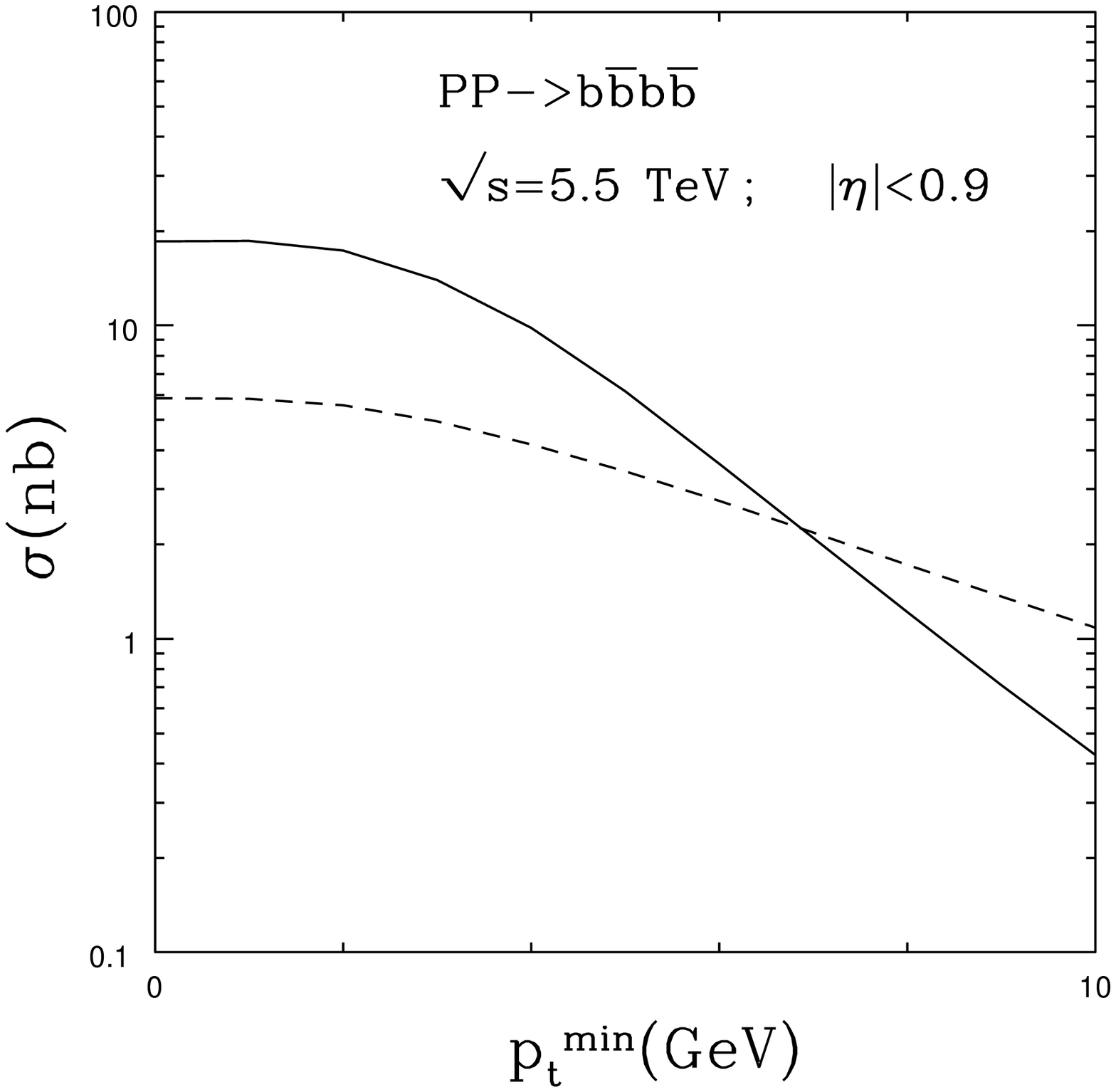,height=3.1in}
\epsfig{figure=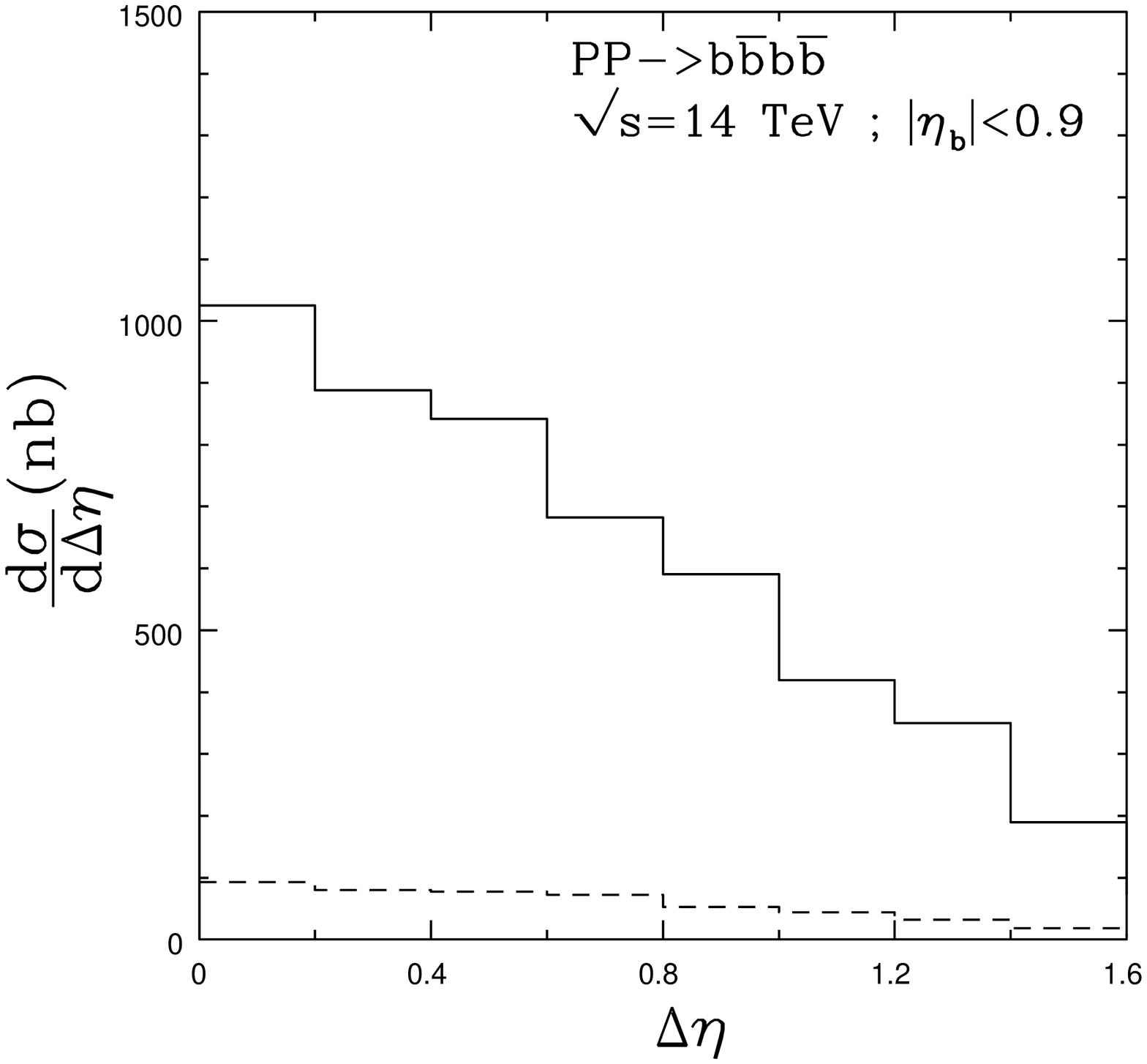,height=3.1in}
\epsfig{figure=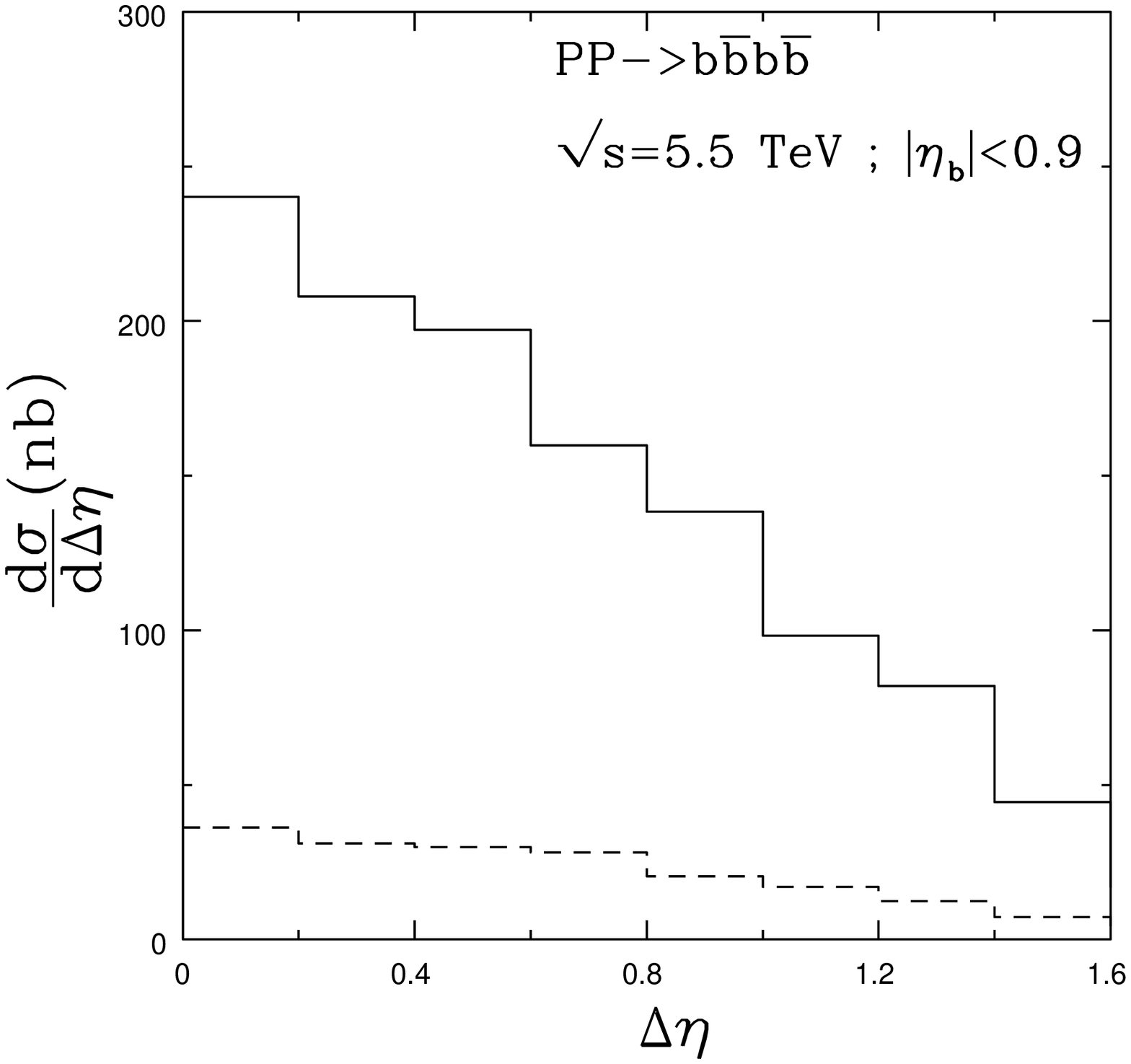,height=3.1in} \caption{Upper
figures: cross sections, for producing two pairs $b\bar b$ with
$|\eta|\le.9$ and with c.m. energies of 14 TeV and 5.5 TeV, as a
function of $p_t^{min}$, the smallest among all the transverse
momenta of the produced quarks. Lower figures:  pseudorapidity
difference distributions  for  two  $b$-quarks with $|\eta|\le.9$
and with c.m. energies of 14 TeV and 5.5 TeV. The continuous lines
are the double parton scattering $(2\to2)^2$, the dashed lines are
the leading QCD $2\to4$.}
\end{center}
\end{figure}

To access the transverse space structure one needs therefore to
measure the different scale factors which characterize the
multiple parton collision processes. By selecting final states as
$jjjj$, $b{\bar b}b{\bar b}$, $c{\bar c}c{\bar c}$ one would
measure $\Theta^{gg}_{gg}$, while final states with a prompt
photon as, $\gamma jjj$, $\gamma jb{\bar b}$ or $\gamma jc{\bar
c}$ will allow one to obtain $\Theta^{qg}_{gg}$ and to test, by
comparing with the CDF result, its energy dependence. The scale
factor $\Theta^{qg}_{qg}$ may be measured by looking at double
parton collisions with a Drell-Yan pair accompanied by two
minijets or by a pair of heavy quarks. Another interesting case is
that of the production of two equal sign $W$ bosons, which, at
relatively low transverse momenta, is dominated by double parton
collisions\cite{Kulesza:1999zh} and which allows one to have
information on the correlation in transverse space of valence
quarks.

All final states produced by double parton collisions can be
produced also by the more conventional leading QCD mechanism,
where the hard process is initiated by two partons. The
contribution of multiparton scatterings is however enhanced with
energy, in such a way that, while the probability of a double
parton collision generating two equal sign $b$-quarks is rather
small at Tevatron energy, one may estimate a cross section of the
order of $1$ $\mu b$ to observe two equal sign $b$ quarks with
$|\eta|\le.9$ at LHC. To compare the two sources of two equal sign
$b$ quark production, in the ALICE detector, the rates of the two
processes are plotted in the fig. 4. The c.m. energies are 14 TeV
and 5.5 TeV. In the upper figures the curves are plotted against
$p_t^{min}$, which is the smallest among all the transverse
momenta of the produced quarks. All the quarks are inside the
pseudorapidity interval $|\eta|\le.9$. The lower figures represent
the distributions of the pseudorapidity difference between  two
$b$-quarks. In this last case the two $b$-quarks only are within
$|\eta|\le.9$.
 All processes have been evaluated at the lowest
order in $\alpha_s$; a factor $k=5.5$ has been however introduced
to account for higher order corrections, which have been assumed
to be the same in the $2\to2$ and in the $2\to4$ processes. The
value of the $k$-factor has been obtained using the results on
heavy quark production in the $k_t$-factorization
approach\cite{Shabelski:2001vd}.
 As a scale factor for the double collision term the
value of $\sigma_{eff}$ measured by CDF has been used. The leading
QCD $2\to4$ contribution is represented by the dashed lines and
the double parton scattering term by the continuous lines. In the
low $p_t^{min}$ region double parton scatterings dominate over the
leading QCD $2\to4$ process by a large factor. Preliminary
estimates indicate that the effect of multiple parton interactions
is even more pronounced when considering $c{\bar c}$ pairs
production. The large values obtained for the double parton
scattering cross sections are an indication that on one hand one
should be able to isolate easily the double parton scattering
contributions while, on the other, there should be a sizable
probability of observing also triple, quadruple etc. parton
collision processes.

As the single scattering term is proportional to the average
number of partonic collisions, the double is proportional to the
dispersion in the distribution of multiple collisions. The
relatively small value measured for $\sigma_{eff}$, whose inverse
is proportional to the double scattering term, is therefore an
indication of large fluctuations in the number of produced jets.
One therefore expects a relatively large number of events without
much activity and a relatively small number of events with many
more large $p_t$ fragments than average. Given the large cross
section expected at the LHC a similar effect should be observed
also in charm production. At the LHC the inclusive cross section
of $c{\bar c}$ production, integrated over all phase space, might
in fact be of the order of $20$ mb, while the corresponding double
parton scattering cross section might reach values as high as $12$
mb. One would then observe a relatively small number of events
with charm, with the unusual feature of being however
characterized by a considerably large number of $c{\bar c}$ pairs.
In such a scenario triple and even quadruple parton collisions
leading to several $c{\bar c}$ pairs would be easily measurable.

The simplest estimate of the triple scattering cross section may
be obtained by neglecting all correlations in $x$ and assuming
that the scale factors do not depend on the different species of
interacting partons. In that case all unknowns of the triple
scattering process are reduced to a single quantity, so that, for
three identical interactions, one may write
\begin{equation}
\sigma_T=\frac{1}{3!}\frac{\sigma_S^3}{\tau\sigma_{eff}^2}
\end{equation}
where $\tau$ is the unknown, dimensionless quantity (of order
unity) to be measured by the triple scattering experiment, while
$\sigma_{eff}$ is the scale factor already measured in a double
parton scattering experiment. More general expressions of the
triple, quadruple etc. cross sections may be easily written
introducing different scale factors, depending on the different
species of interacting partons and expressed through integrals of
the n-parton densities in transverse space, in a way analogous to
the double scattering cross section in Eq.s
(\ref{sigmad1})(\ref{fb}).

\section{Summary}

All multiple parton scattering effects increase considerably at
the LHC, enhancing features of the typical inelastic event as the
increase of $\langle p_t\rangle$ with multiplicity, already
observed by UA1, and changing the distribution in multiplicity of
the jets produced with a relatively low $p_t$. A characteristic
feature of multiparton interactions is the dependence of the cross
section on dimensional scale factors, whose origin is geometrical
and which are introduced into dynamics by the non-perturbative
input to the process. Differently with respect to the usual
non-perturbative input to a large $p_t$ - fixed $x$ interaction,
the non-perturbative input to a multiparton process is in fact a
multiparton distribution, which is a dimensional quantity and
contains information on the non perturbative structure independent
on the knowledge of the one-body structure functions usually
investigated in large $p_t$ processes. In the simplest case of a
double collision the information is represented by the factors
$\Theta^{ij}_{kl}$ discussed above. It should be stressed that one
cannot obtain the factors $\Theta^{ij}_{kl}$ by testing the hadron
with an elementary probe. Multiparton scatterings are hence a
unique tool for a deeper investigation on the hadron structure,
which have the potential of providing a rather interesting new
physical output. The scale factors are in fact a measure of the
typical transverse distance between different pairs of partons
inside the hadron, so their determination gives one access to the
three-dimensional structure of the hadron.

\end{document}